\documentclass[12pt]{article}
\usepackage{graphicx}
\usepackage{float} 

\begin{document}
\begin{titlepage}

\title{Kinetic Energy and Angular Momentum of Free Particles in the
Gyratonic pp-Waves Space-times}

\author{J. W. Maluf$\,^{(1)}$, J. F. da Rocha-Neto$\,^{(2)}$, \\
S. C. Ulhoa$\,^{(3)}$, and F. L. Carneiro$\,^{(4)}$ \\
Instituto de F\'{\i}sica, 
Universidade de Bras\'{\i}lia\\
70.919-970 Bras\'{\i}lia DF, Brazil\\}
\maketitle
\bigskip
\bigskip

\begin{abstract}
Gyratonic pp-waves are exact solutions of Einstein's equations that represent
non-linear gravitational waves endowed with angular momentum. We 
consider gyratonic pp-waves that travel in the $z$ direction and  whose time
dependence on the variable $u={1 \over \sqrt{2}}(z-t)$ is given by gaussians,
so that the waves represent short bursts of gravitational radiation propagating
in the $z$ direction. We evaluate numerically the geodesics and velocities of 
free particles in the space-time of these waves, and find that after the passage
of the waves both the kinetic energy and the angular momentum per unit mass of 
the particles are changed. Therefore there is a transfer of energy and angular 
momentum between the gravitational field and the free particles, so that the 
final values of the energy and angular momentum of the free particles may be 
smaller or larger in magnitude than the initial values.
\end{abstract}
\thispagestyle{empty}
\vfill
\par

\bigskip
{\footnotesize
\noindent (1) wadih@unb.br, jwmaluf@gmail.com\par
\noindent (2) rocha@fis.unb.br\par
\noindent (3) sc.ulhoa@gmail.com\par
\noindent (4) fernandolessa45@gmail.com\par}

\end{titlepage}
\newpage

\section{Introduction}
The field equations of the theory of general relativity admit a class of exact
solutions that represent non-linear 
plane gravitational waves. These solutions are known
as the Kundt family of space-times \cite{Kramer}, and were thoroughly studied 
by Ehlers and Kundt \cite{Ehlers,Ehlers-2}. They may describe 
short bursts of gravitational radiation. The idea is that far from the
source we may approximate a realistic gravitational wave by an exact plane wave.
Some of these non-linear plane waves are vacuum solutions of Einstein's 
equations, and are idealized manifestations of the gravitational
field, in similarity to the source-free wave solutions of Maxwell's equations,
which are excellent descriptions of realistic monochromatic plane
electromagnetic waves. Ehlers and Kundt \cite{Ehlers,Ehlers-2} asserted that 
the vacuum pp-waves (parallelly propagated plane-fronted waves) are geodesicaly
complete (see Ref. \cite{Flores}), and also showed that one may add the 
amplitudes of two pp-waves running in the same direction, in which case the 
principle of linear superposition holds (section 2-5.3 of Ref. \cite{Ehlers}).
They argue that this fact indicates a striking analogy with Maxwell's theory.

Penrose \cite{Penrose} also 
investigated these waves, and concluded that ``{\it it is fair to assume that
they} (the non-linear gravitational waves) {\it are no less physical (as 
idealizations) than their flat space couterparts}'', although he also concluded
that is not possible to imbed these waves globally in any hyperbolic 
pseudo-Euclidean space. The difficulty, according to Penrose, is that the total 
(gravitational) energy of such plane waves is infinite. However, the same 
feature takes place for idealized waves in flat space. It must be noted that the
pp-waves do not imply or lead to the violation of any physical principle or 
physical property. They are ordinary manifestations of the theory.

Gyratonic pp-waves represent the exterior gravitational field of a spinning 
source that propagates at the speed of light. The exterior space-time of a 
localized spinning source, that moves at the speed of light, was first 
studied by Bonnor \cite{Bonnor}, later by Griffiths \cite{Griffiths}, and 
rediscovered by Frolov and collaborators 
\cite{Frolov-1,Frolov-2,Frolov-3,Frolov-4}, who attempted to relate this 
space-time with the model of spinning particles, the ``gyratons''. The gyratonic
pp-waves were recently investigated by Podolsk\'{y}, Steinbauer and \v{S}varc 
\cite{Podolsky}. These authors studied several properties of the non-linear 
waves, such as the impulsive limit, the geodesic
motion, and also emphasized the role of the off-diagonal components in the line
element, that is described in terms of Brinkmann coordinates \cite{Brinkmann}.
The gyratonic pp-waves are non-linear gravitational waves endowed with angular 
momentum. 

In this article we evaluate the geodesics and velocities of free particles in
the space-time of gyratonic pp-waves, by means of a numerical analysis, 
and show that there is not only a transfer of energy between the free particles
and the gravitational wave \cite{arxiv2017},
but also a transfer of angular momentum. Both transfers of energy and angular
momentum between the particles and the field take place locally, i.e. in the 
regions where the particles are located. This is an argument in favour of
the localization of gravitational energy. It does not seem reasonable to 
consider that a local variation of energy of a free particle is related to the
instantaneous variation of the total gravitational energy, evaluated at 
spacelike (or null) infinity, for instance. We note that Ehlers and Kundt 
already investigated the action of a pp-wave on free particles initially at
rest, and concluded that after the passage of the wave, the particles acquire
velocities, both transverse and longitudinal to the wave (see section 2-5.8 of
\cite{Ehlers}). We show in this article, as in Ref. \cite{arxiv2017},  
that the final energy of the particles may be smaller than the initial energy.

In section 2 we describe the gyratonic pp-waves, according to the presentation 
of Ref. \cite{Podolsky}. The graphic results are presented in section 3, and the
final comments in section 4.

\section{Gyratonic pp-waves}

The study of the gyratonic pp-waves emerged from the investigation by 
Bonnor \cite{Bonnor} and by Griffiths \cite{Griffiths} of the interior and 
exterior fields of a 
spinning ``null fluid'', which is a configuration that travels at the speed of
light. A review of this issue is given in section 18.5 of Ref. \cite{GP}. In the
following we will adopt the presentation of Ref. \cite{Podolsky}. We will 
consider a wave that travels along the $z$ direction. The gyratonic source 
$(\varrho, j_i)$ is localized along the axis $x=y=0$, or in the immediate 
neighbourhood around this axis. The energy-momentum of the source is prescribed 
by the radiation density $T_{uu}=\varrho$, and by the terms $T_{ui}=j_i$ that 
represent the spinning character of the source.
The line element is given in terms of Brinkmann coordinates 
\cite{Brinkmann}, with an off-diagonal term. It is given by \cite{Podolsky}

\begin{equation}
ds^{2}=d\rho^{2}+\rho^{2}d\phi^{2}+2\,dudv-2J\,dud\phi+H\,du^{2}\,.
\label{1}
\end{equation}

\noindent The integration of the field equations in the vacuum region, outside 
the matter source, imply that the functions
$H=H(u,\rho,\phi)$ and $J=J(u,\rho,\phi)$ are periodic in $\phi$. The general 
form of these functions is \cite{Podolsky}

\begin{equation}
J=\omega(u)\rho^2+\chi(u,\phi)\,,
\label{2}
\end{equation}

\begin{equation}
H=\omega^2(u)\rho^2+2\omega(u)\chi(u,\phi)+H_0(u,\rho,\phi)\,,
\label{3}
\end{equation}
where $\omega(u)$ and $\chi(u,\phi)$ are arbitrary functions of $u$ and 
$(u,\phi)$, respectively. If $\chi$ is taken to be independent of the angular
coordinate $\phi$, i.e., if $\chi=\chi(u)$, then the function $H_0$ must only
satisfy the equation 

\begin{equation}
\nabla^2 H_0=\biggl( {{\partial^2} \over{\partial x^2}}+ 
{{\partial^2} \over{\partial y^2}} \biggr)H_0=0\,.
\label{4}
\end{equation}
We will adopt this simplification in the considerations below.

In flat space-time, the variables $u$ and $v$ are related to the ordinary time 
variable $t$  and to the usual cylindrical coordinate $z$ according to 

\begin{equation}
u = {1\over {\sqrt{2}}}(z-t)\,,
\label{5}
\end{equation}

\begin{equation}
v = {1\over {\sqrt{2}}}(z+t)\,.
\label{6}
\end{equation}
We use the relations above to carry out a coordinate transformation of the 
metric tensor, from the $(u,\rho,\phi,v)$ coordinates to $(u,\rho,\phi,z)$ 
coordinates, so that the line element is rewritten as 

\begin{equation}
ds^{2}=(H-2)du^{2}+d\rho^{2}+\rho^{2}d\phi^2 + 2 \sqrt{2}\,dudz-2J\,dud\phi\,.
\label{7}
\end{equation}
The flat space-time is obtained if we make $H=J=0$.
From the metric tensor above we obtain, by means of the standard procedure, the
Lagrangian for a free particle of mass $m$ that travels along geodesics 
labelled by an affine parameter $\lambda$. The Lagrangian is given by 

\begin{equation}
L= \frac{m}{2}\left[(H-2)\dot{u}^{2}+\dot{\rho}^{2}+\rho^{2}\dot{\phi}^{2}
+\sqrt{8}\,\dot{u}\dot{z}-2J\,\dot{u}\dot{\phi}\right]\,,
\label{8}
\end{equation}
where the dot represents variation with respect to $\lambda$.

By solving the equations of motion we find, in similarity to the analysis of 
Ref. \cite{arxiv2017}, that ${d\,}^2u/d\lambda^2=0$. As a consequence we may set
$\dot{u}=1$, and thus we take the coordinate $u$ as the affine parameter along
the geodesic. In this way, the set of geodesic equations read

\begin{eqnarray}
\ddot{\rho} + \partial_{\rho}J\dot{\phi} - 
\rho \dot{\phi}^{2}-\frac{1}{2} \partial_{\rho}H&=&0 \,,  \label{9} \\
\ddot{\phi} + \frac{2}{\rho}\dot{\rho}\dot{\phi}-
\frac{\partial_{\rho}J}{\rho^{2}}\dot{\rho}-
{1\over {2\rho^2}} \biggr[2\partial_{u}J
+\partial_{\phi}H \biggl]&=&0 \,, \label{10} \\
\ddot{z} + {1\over{\sqrt{2}\rho^{2}}}\biggr[\rho^{2}\partial_{\rho}H
-J\partial_{\rho}J\biggl]\dot{\rho}
-\frac{\partial_{\phi}J}{\sqrt{2}}\dot{\phi}^2+
\frac{\partial_{\phi}H}{\sqrt{2}}\dot{\phi}&{}& \nonumber \\
+{1\over {\sqrt{2}\rho}}\biggl[{2J-\rho
\partial_{\rho}J} \biggr]   \dot{\rho}\dot{\phi} 
+{1\over {\sqrt{8}\rho^2}}\biggl[{\rho^{2}\partial_{u}H
-J(\partial_{\phi}H+2\partial_{u}J)}\biggr]&=&0\,.\label{11}
\end{eqnarray}
These equations are equivalent to the geodesic equations given by Eq. (75) of
Ref. \cite{Podolsky}, presented in the $(\rho,\phi,v)$ coordinates.

The function $\omega(u)$ is related to a rigid rotation of the space-time, and
can be removed by a gauge transformation \cite{Podolsky}. Without loss of 
generality, we can make $\omega=0$. Therefore, the remaining arbitrary functions
are

\begin{equation}
J=\chi(u)\,,
\label{12}
\end{equation}

\begin{equation}
H=H_0(u,\rho,\phi)\,.
\label{13}
\end{equation}
The $u$-dependence of these two functions will be given by gaussians, that 
represent short bursts of gravitational waves. We already
discussed in Ref. \cite{arxiv2017} that in order to obtain qualitative features
of the geodesic equations, it makes no difference whether we use gaussians or
derivatives of gaussians. We note that in the analysis of the gravitational 
memory effect carried out in Ref. \cite{ZDGH2}, use
is made of the first, second and third derivatives of gaussians. The advantage 
of using the second derivative is that it dispenses a multiplicative dimensional
constant in the metric tensor components \cite{arxiv2017}. By requiring $J=0$,
we obtain the pp-wave considered in Refs. \cite{arxiv2017,ZDGH2}.

As for the $(x,y)$ or $(\rho,\phi)$ (where $x=\rho\cos\phi$ and 
$y=\rho\sin\phi$) dependence of the function $H_0$, that must
satisfy Eq. (\ref{4}), we will establish two possibilities. The first one is the
$\times$ polarization. It reads

\begin{equation} 
H_{0}=\frac{1}{2}(xy)\frac{d\,^{2}}{du^{2}}(e^{-u^{2}/\sigma^2})\,,
\label{14}
\end{equation}
and the function $\chi(u)$ is chosen to be

\begin{equation} 
\chi={1\over {10}} e^{-u^{2}/\sigma^2}\,.
\label{15}
\end{equation}
which also represents a burst. In the expressions above, $\sigma$ is a constant
with dimension of length. The multiplicative factor $1/10$ is introduced 
only for graphical reasons (for adjusting the figures).
For our purposes, it makes no qualitative difference whether we use the $\times$ 
polarization, or the $+$ polarization given by  $x^2-y^2$, instead of 
$xy$. 

The second choice of $H_0$ is

\begin{equation} 
H_{0}=\frac{1}{4}\lbrack \log (\rho/\rho_0)\rbrack  
\frac{d\,^{2}}{du^{2}}(e^{-u^{2}/\sigma^2})\,,
\label{16}
\end{equation}
and the $\chi$ function is given by

\begin{equation} 
\chi= {1\over {10}} e^{-(0.1\,u^2/\sigma^2)}\,.
\label{17}
\end{equation}
In Eqs. (\ref{16}) and (\ref{17}), not only $\sigma$, but also $\rho_0$ is a
constant with dimension of length in natural units. For simplicity, both 
constants will be taken to be $\sigma=\rho_0=1$. Except for the second
derivative of the gaussian, the function in Eq. (\ref{16}) corresponds to the
Aichelburg-Sexl type monopole \cite{Aichelburg}.

\section{The geodesic of free particles}

In this section we present the graphical results obtained by solving the 
geodesic equations (\ref{9}), (\ref{10}) and (\ref{11}) numerically, by means of
the program MATHEMATICA. We also plot the Kinetic energy per unit mass $K$
of the free particles before and after the passage of the wave (in which case 
the space-time is flat), as a function of
the variable $u$. In view of Eq. (\ref{5}), $K$ is obtained from the equation 

\begin{equation}
4K=\dot{\rho}^{2} + \rho^{2}\dot{\phi}^{2} + \dot{z}^{2}\,,
\label{18}
\end{equation}
where the dot represents variation with respect to $u$ (we would have $2K$ on 
the left hand side, if the variation were with respect to $t$).

\subsection{$H_0$ given by equation (\ref{14})}

\begin{figure}[!htb]
   \begin{minipage}{0.49\linewidth}
     \centering
     \includegraphics[width=\textwidth]{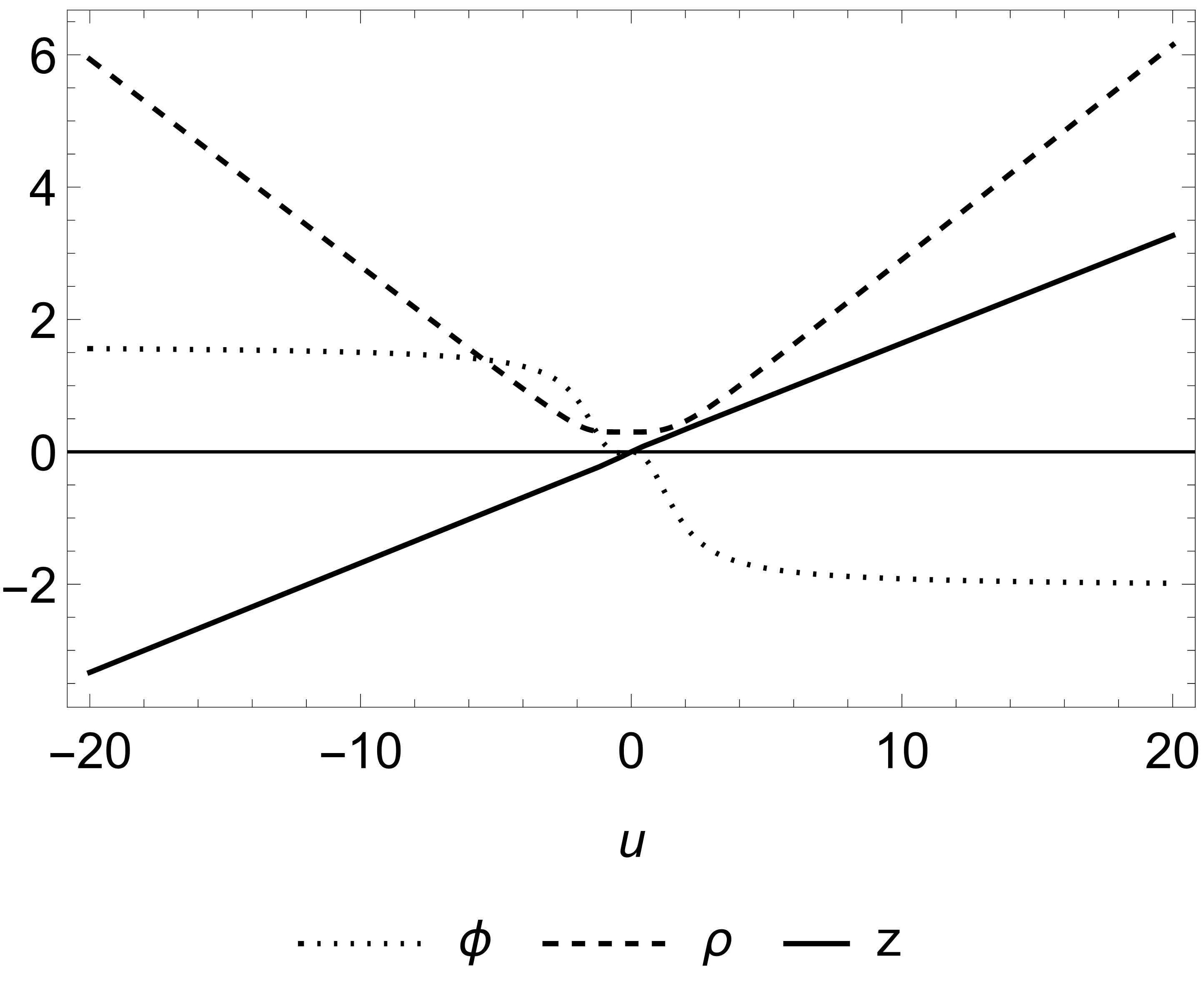}
     \caption{Coordinates when $H_0$ is given by (\ref{14}) and for the 
     initial conditions (\ref{19}).}\label{Figure1}
   \end{minipage}\hfill
   \begin {minipage}{0.49\linewidth}
     \centering
     \includegraphics[width=\textwidth]{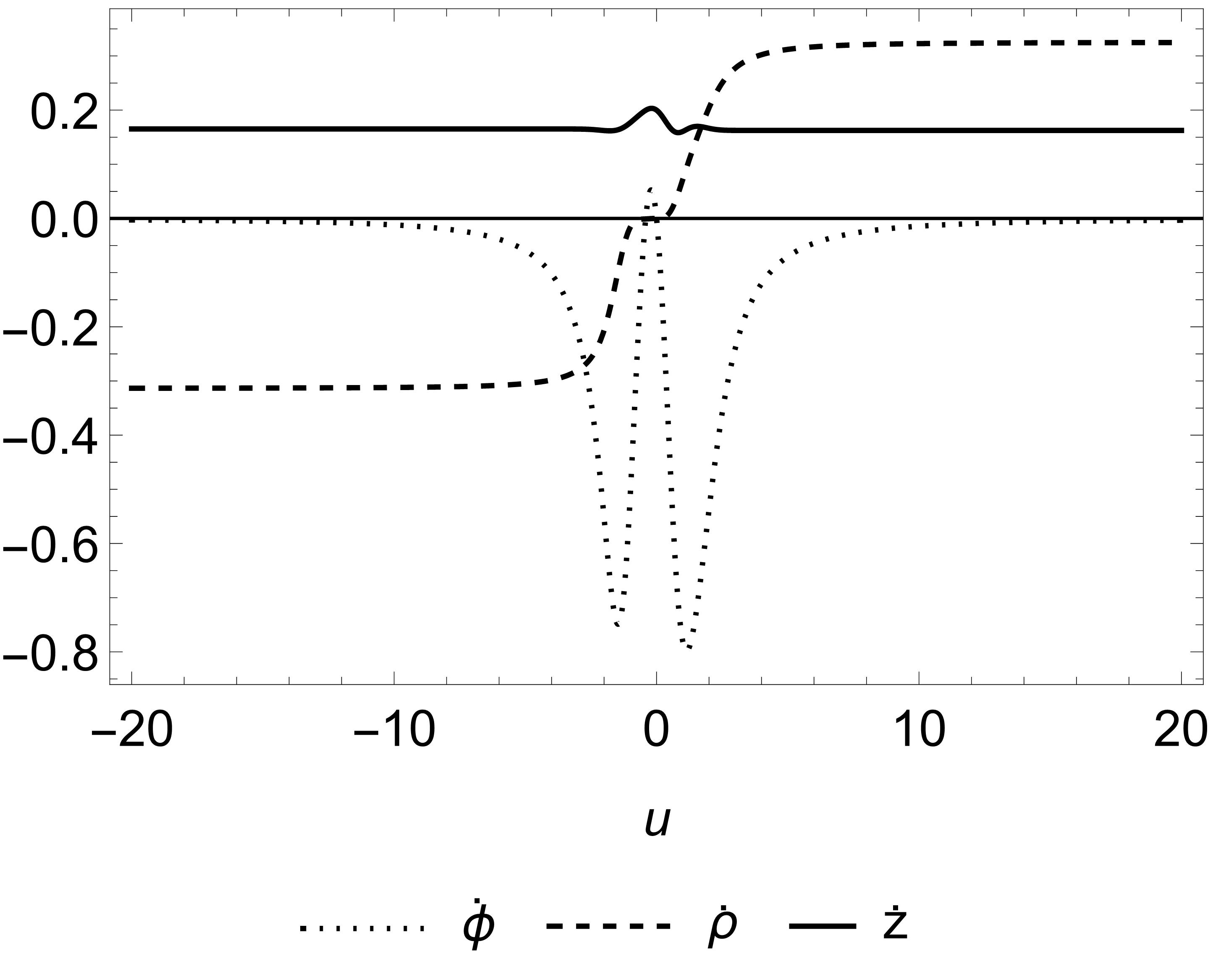}
     \caption{Velocities when $H_0$ is given by (\ref{14}) and for the 
     initial conditions (\ref{19}).}\label{Figure2}
   \end{minipage}
	\end{figure}

For the initial conditions given by

\begin{equation} \label{eq:ini1}
\rho_{0}=0.3, \ \ \ \phi_{0}=0, \ \ \ z_{0}=0, \ \ \ \dot{\rho_{0}}=0,
\ \ \ \dot{\phi_{0}}=0,\ \ \ \dot{z_{0}}=0.2\,,
\label{19}
\end{equation}
for $u=0$, 
we obtain the geodesic behaviour displayed in Figures [1] and [2], where both
coordinates (Figure [\ref{Figure1}]) and velocities (Figure [\ref{Figure2}])
are given as functions of $u$.

The trajectory of the free particle in the $x-y$ plane is plotted in 
Figure [3]. The initial state of the particle is in the upper part of the Figure.
The particle acquires angular momentum during the passage of the gravitational
wave, and then resumes the inertial movement along a straight line in the lower
part of the Figure. It is clear that there is a transfer of angular momentum
between the wave and the free particle. An analogous behaviour has been studied
also in Ref. \cite{ZDGH2}, but in the context of a pp-wave with $J=0$, with
the polarization $x^2-y^2$.

\begin{figure}
	\centering
		\includegraphics[width=0.17\textwidth]{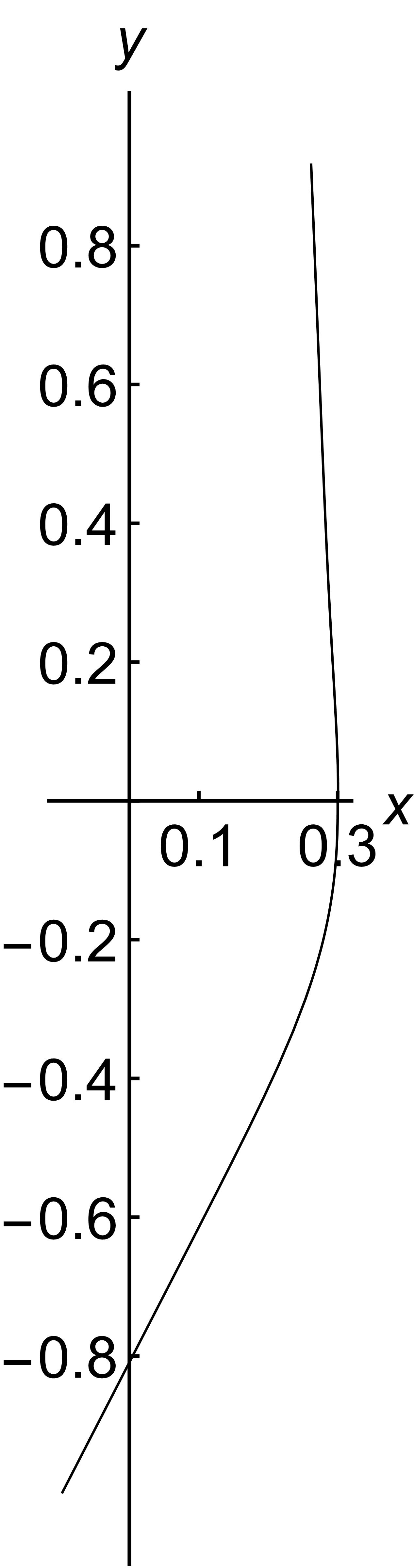}
	\caption{Trajectory in the $x-y$ plane when $H_0$ is given by (\ref{14})
	and for the initial conditions (\ref{19}).}
	\label{Figure3}
	\end{figure}

\begin{figure}[!htb]
   \begin{minipage}{0.49\linewidth}
     \centering
     \includegraphics[width=\textwidth]{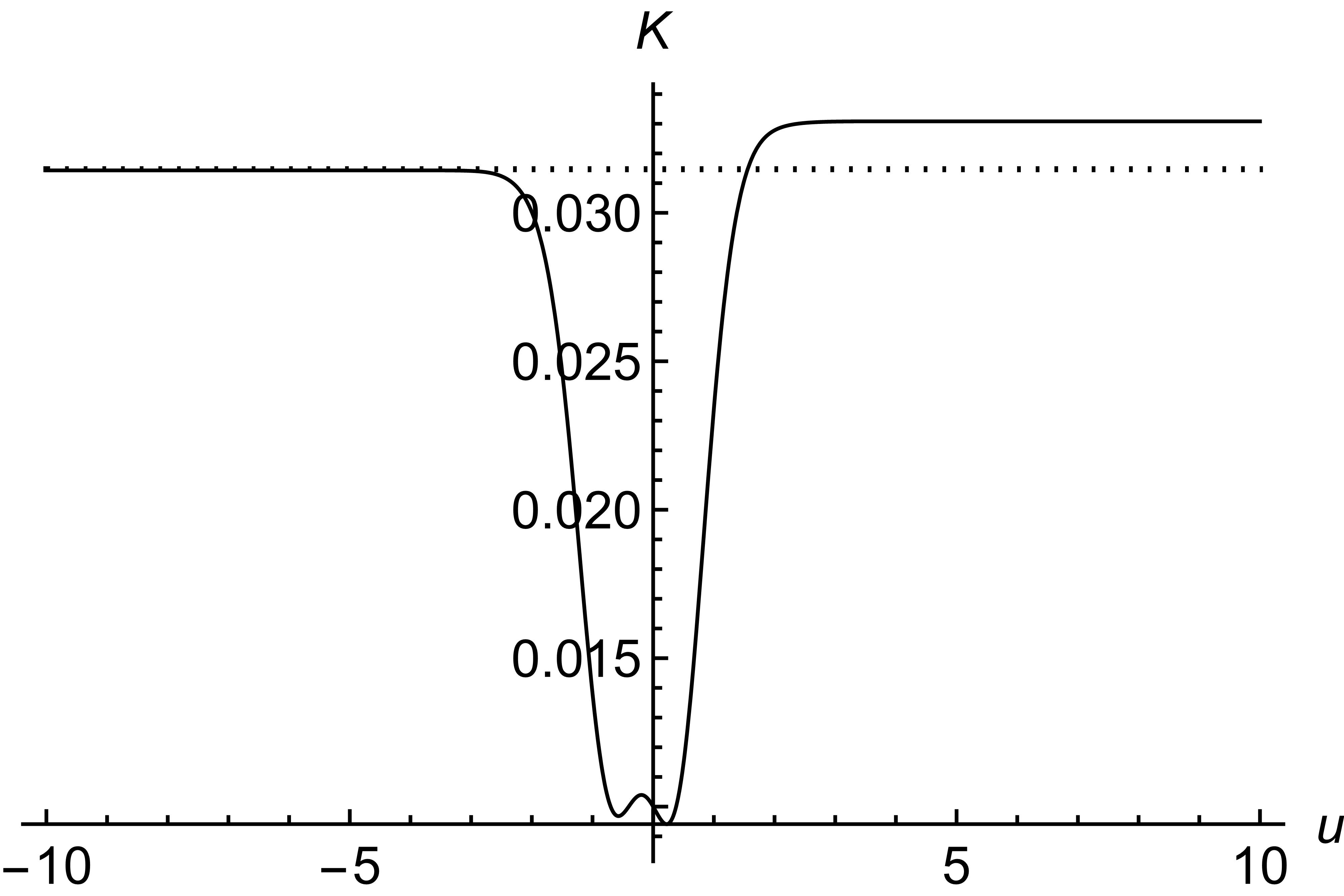}
     \caption{Kinetic energy when $H_0$ is given by (\ref{14}) and initial 
     	conditions (\ref{19}).}\label{Figure4}
   \end{minipage}\hfill
   \begin {minipage}{0.49\linewidth}
     \centering
     \includegraphics[width=\textwidth]{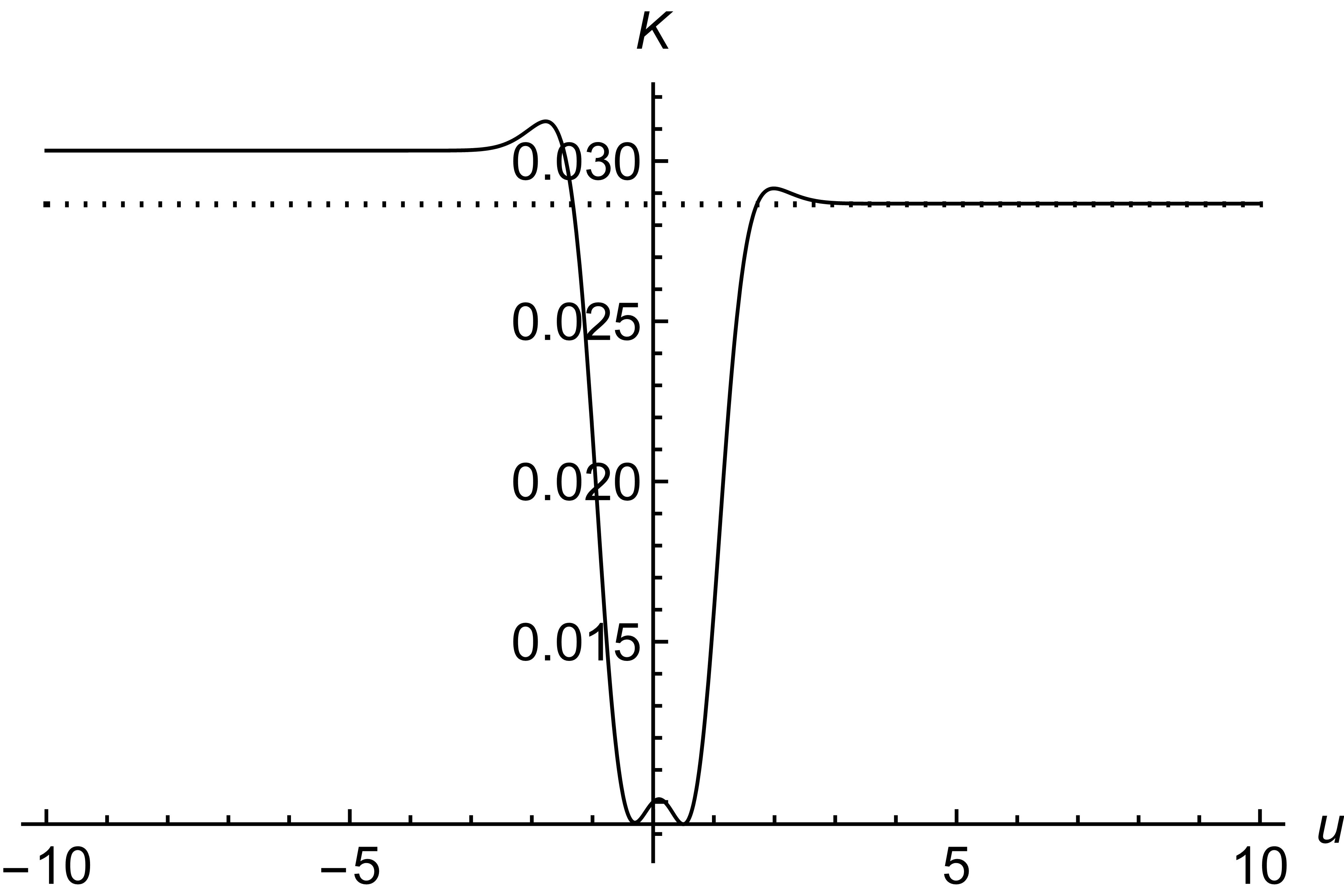}
     \caption{Kinetic energy when $H_0$ is given by (\ref{14}) and initial 
     	conditions (\ref{20}).}\label{Figure5}
   \end{minipage}
	\end{figure}

We also choose an alternative set of initial conditions, 

\begin{equation} 
\rho_{0}=0.3, \ \ \ \phi_{0}=\frac{\pi}{3},\ \ \ z_{0}=0,\ \ \ \dot{\rho_{0}}=0,
\ \ \ \dot{\phi_{0}}=0, \ \ \ \dot{z_{0}}=0.2 \,,
\label{20}
\end{equation}
also for $u=0$, 
where the only difference with respect to initial conditions (19) is the value 
of the initial angle $\phi_0$. The reason for choosing this alternative set 
is to show that the behaviour of the Kinetic energy per unit mass of the
free particle is highly sensitive to the initial conditions. Figure 
[\ref{Figure4}] displays the behaviour of the Kinetic energy by choosing the 
polarization given by Eq. (\ref{14}), with the initial conditions (\ref{19}).
In Figure [\ref{Figure5}] we see the behaviour of the Kinetic energy of the 
particle, after the passage of the wave, when the initial conditions are given 
by Eq. (\ref{20}).

In view of Eq. (\ref{5}), a positive variation of $u$ corresponds to a negative 
variation of the time coordinate $t$. Therefore, if $u$ grows from left to right
in all Figures, the time coordinate $t$ grows from right to left. Thus, we see 
that in Figure [\ref{Figure4}] the final Kinetic energy is smaller than the 
initial energy, whereas in Figure [\ref{Figure5}] the final Kinetic energy is 
larger than the initial energy. This is exactly the behaviour discussed in 
Ref. \cite{arxiv2017}. The passage of a gravitational wave may either increase 
or decrease the energy of a free particle, and therefore there is a local 
transfer of energy between the particle and the wave.

\begin{figure}[hp]
	\centering
		\includegraphics[width=0.60\textwidth]{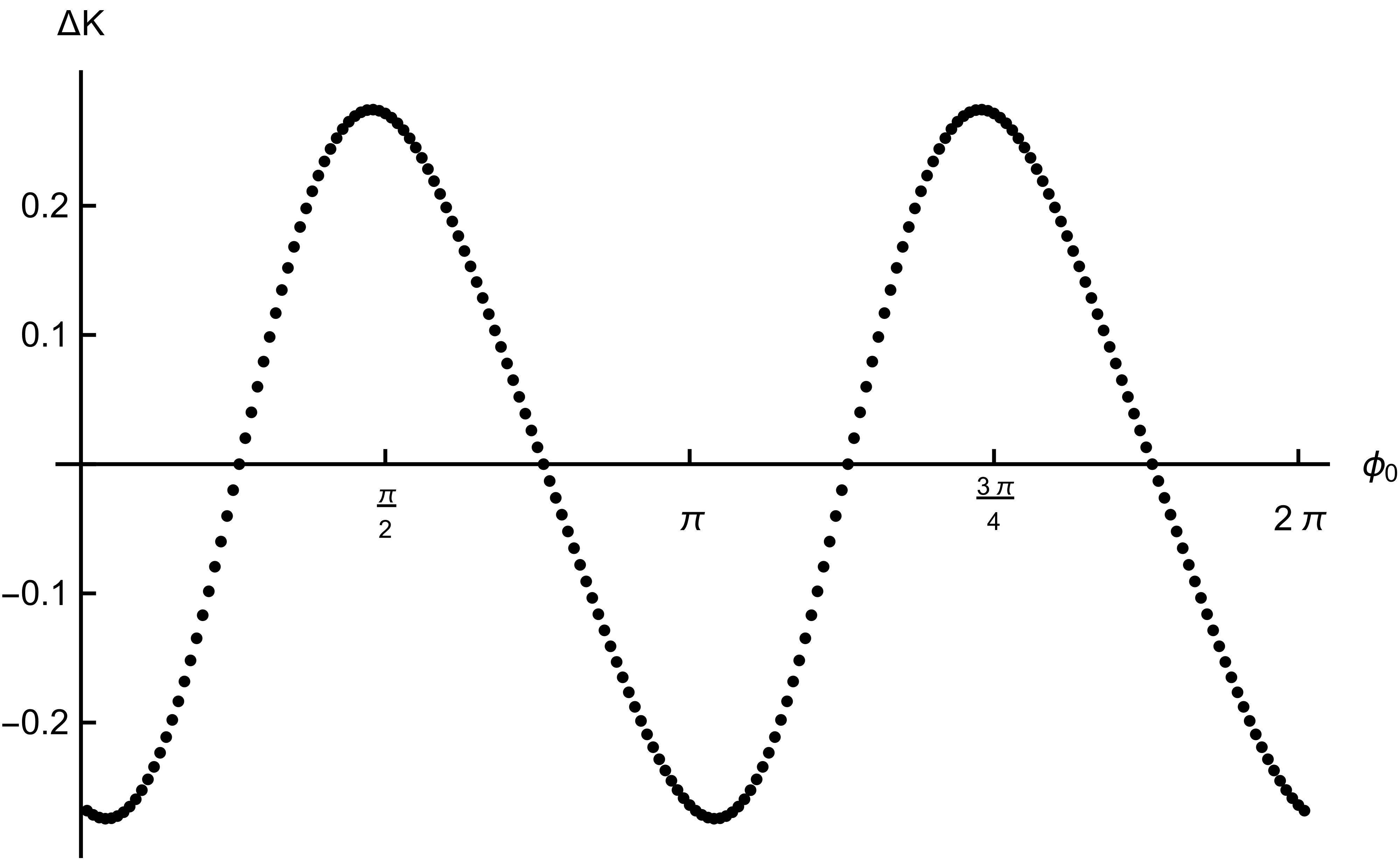}
	\caption{Variation of the Kinetic energy with respect to $\phi_0$, in the 
	initial conditions $(\ref{22})$.}
	\label{deltaK}
\end{figure}

In order to explore a little further the variation of the Kinetic energy, before
and after the passage of the wave (but without the purpose of exhausting the
investigation), we define the quantity

\begin{equation}
\Delta K={{K_f-K_i}\over {K_f+K_i}}\,,
\label{21}
\end{equation}
where $K_f$ is the final Kinetic energy evaluated at $u=-5$, and $K_i$ is the
initial Kinetic energy evaluated at $u=5$ ($\Delta t>0$, since $\Delta u <0$).
Thus, the particle acquires or looses energy according to $\Delta K>0$ or 
$\Delta K<0$, respectively. Note that the gravitational wave considered here is
not axially symmetric. Choosing the initial conditions (always for $u=0$),

\begin{equation} 
\rho_{0}=0.1,  \ \ \ z_{0}=0, \ \ \ \dot{\rho_{0}}=0, 
\ \ \ \dot{\phi_{0}}=0, \ \ \ \dot{z_{0}}=0, 
\label{22}
\end{equation}
we obtain the behaviour displayed by Figure [\ref{deltaK}], where $\Delta K$
varies with respect to $\phi_0$. It is possible to show that by increasing or
decreasing the value of $ \dot{\rho_{0}}$ around $ \dot{\rho_{0}}=0$, the curve
moves as a whole up or down in the Figure, respectively.

\subsection{$H_0$ given by equation (\ref{16})}

\begin{figure}[!htb]
   \begin{minipage}{0.49\linewidth}
     \centering
     \includegraphics[width=\textwidth]{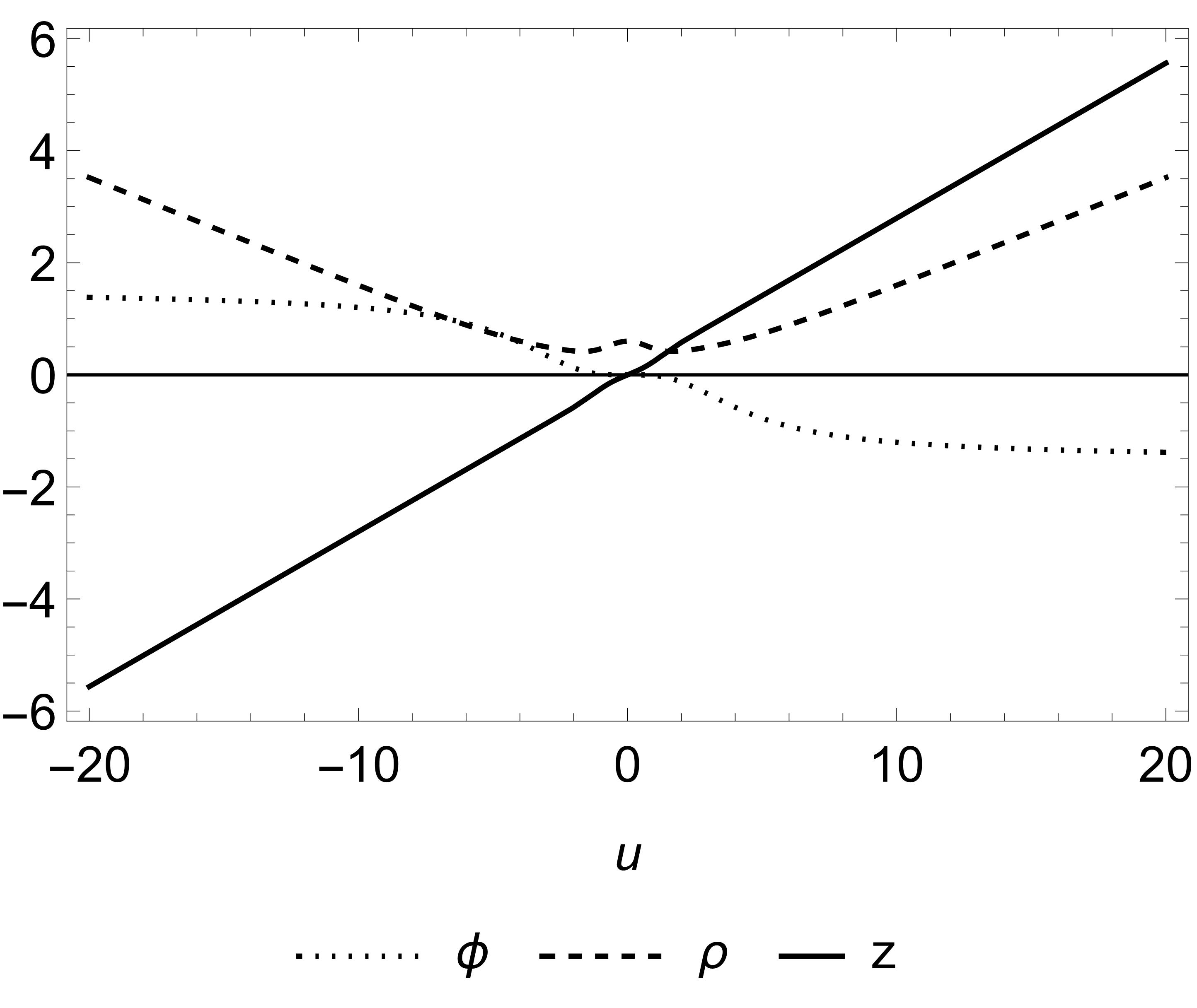}
     \caption{Coordinates when $H_0$ is given by (\ref{16}) and for the 
     initial conditions (\ref{23}).}\label{Figure7}
   \end{minipage}\hfill
   \begin {minipage}{0.49\linewidth}
     \centering
     \includegraphics[width=\textwidth]{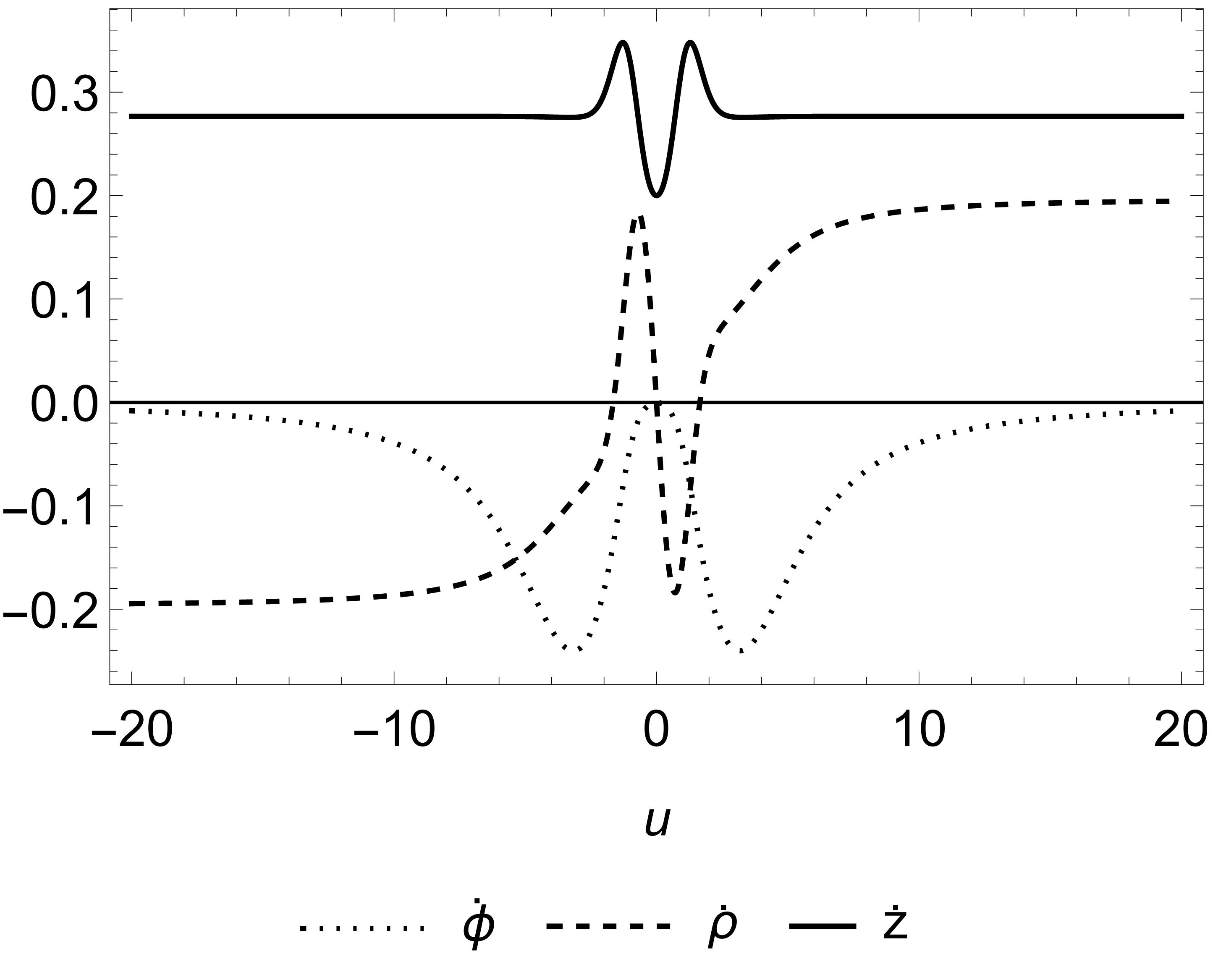}
     \caption{Velocities when $H_0$ is given by (\ref{16}) and for the 
     initial conditions (\ref{23}).}\label{Figure8}
   \end{minipage}
	\end{figure}

Let us now consider the second choice for the function $H_0$, given by Eq. 
(\ref{16}), together with $\chi(u)$ given by (\ref{17}). In this case, the wave
is axially symmetric. The initial conditions are chosen to be

\begin{equation} 
\rho_{0}=0.6, \ \ \ \phi_{0}=0, \ \ \ z_{0}=0, \ \ \ \dot{\rho_{0}}=0,
\ \ \ \dot{\phi_{0}}=0, \ \ \ \dot{z_{0}}=0.2\,.
\label{23}
\end{equation}
The geodesic behaviour and the velocities of a free particle under the action
of the gravitational wave, for the initial conditions above, are displayed in 
Figures [\ref{Figure7}] and [\ref{Figure8}]. 

\begin{figure}
	\centering
		\includegraphics[width=0.32\textwidth]{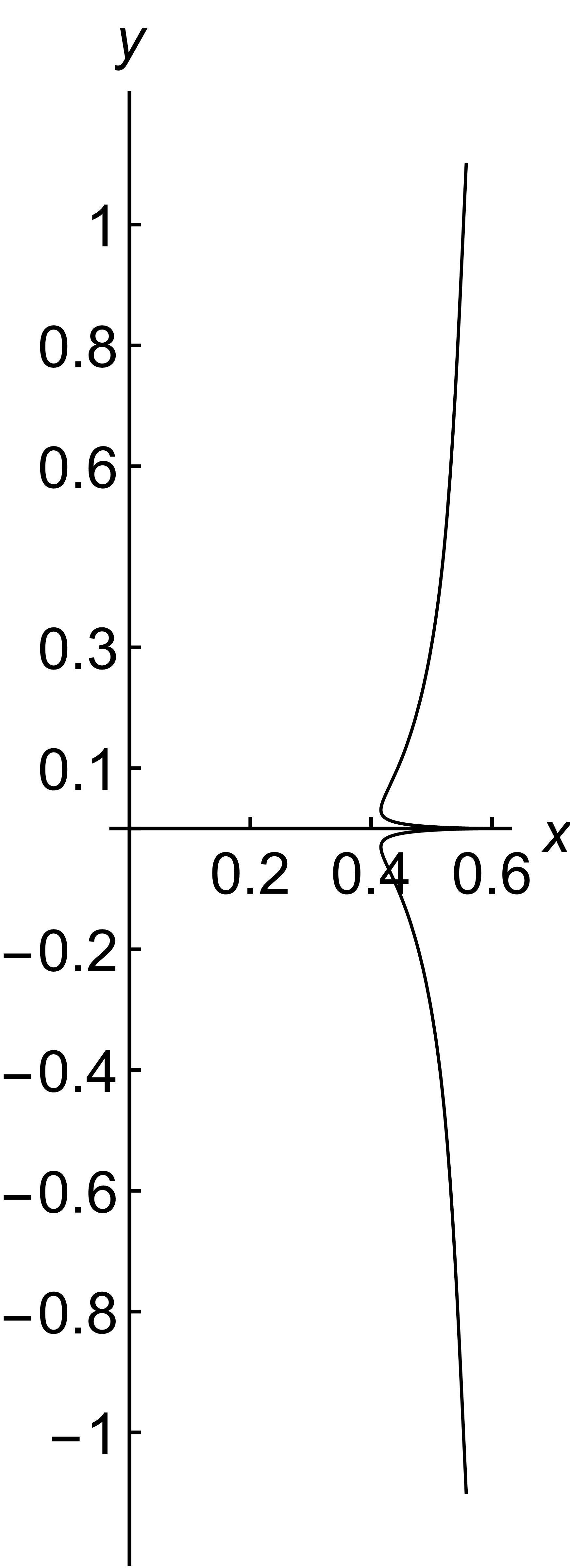}
	\caption{Trajectory in the $x-y$ plane when $H_0$ is given by (\ref{16})
	and for the initial conditions (\ref{23}).}
	\label{Figure9}
	\end{figure}

The trajectory of a free particle in the $x-y$ plane is plotted in Figure 
[\ref{Figure9}]. Similarly to the situation displayed in Figure [\ref{Figure3}], 
the particle acquires angular momentum during the passage of the gravitational 
wave, and then resumes the inertial movement along a straight line. Both the 
initial and final trajectories of the particle are linear. These trajectories
are in the upper and lower part of the Figure, respectively, and are almost 
vertical trajectories. In these regions, the particle is free and away of the 
gravitational field of the wave.

\begin{figure}[!htb]
   \begin{minipage}{0.49\linewidth}
     \centering
     \includegraphics[width=\textwidth]{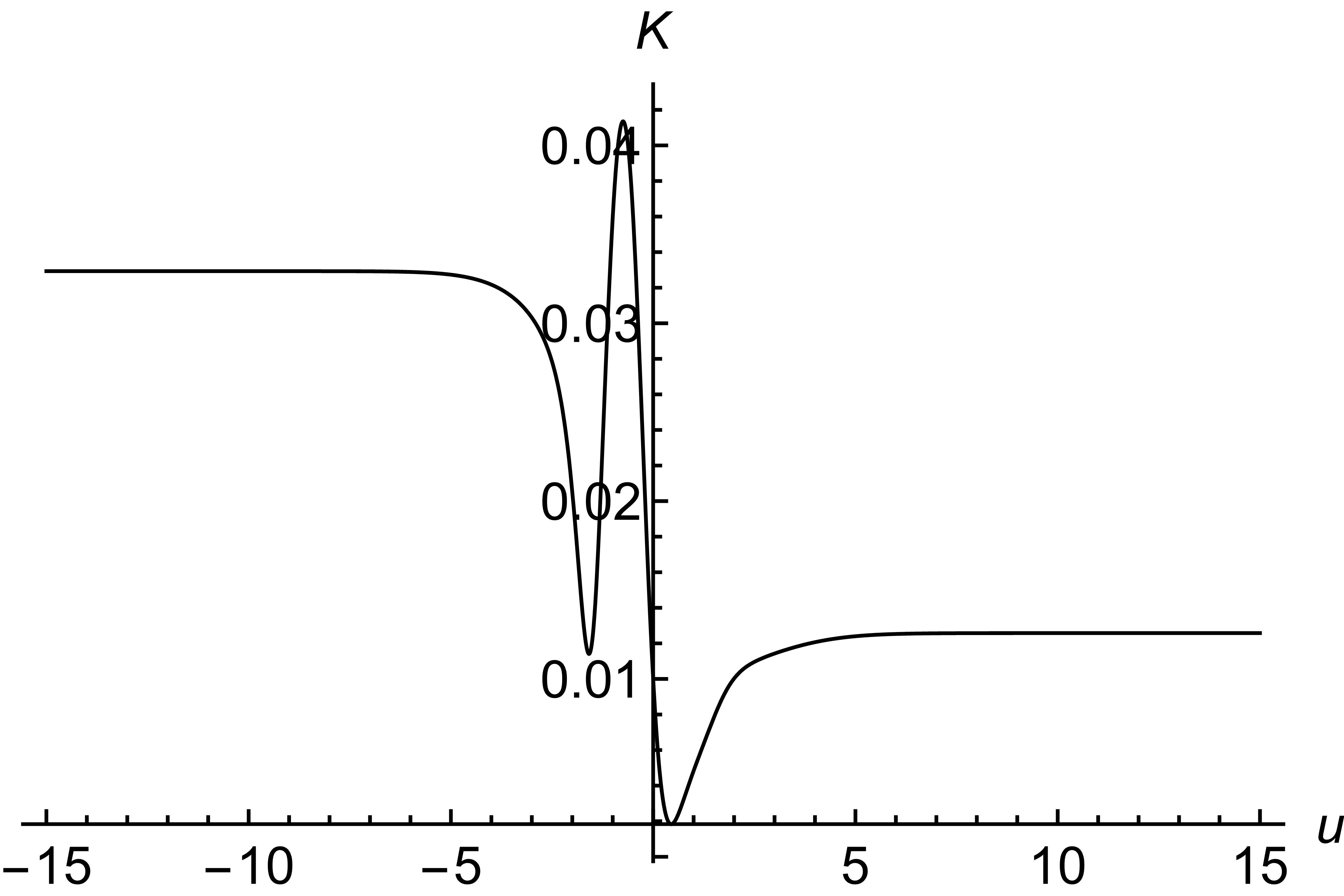}
     \caption{Increasing of the Kinetic energy for the initial conditions 
          (\ref{24}), and for positive sign of $\dot{\rho_{0}}$.}
     \label{Figure10}
   \end{minipage}\hfill
   \begin {minipage}{0.49\linewidth}
     \centering
     \includegraphics[width=\textwidth]{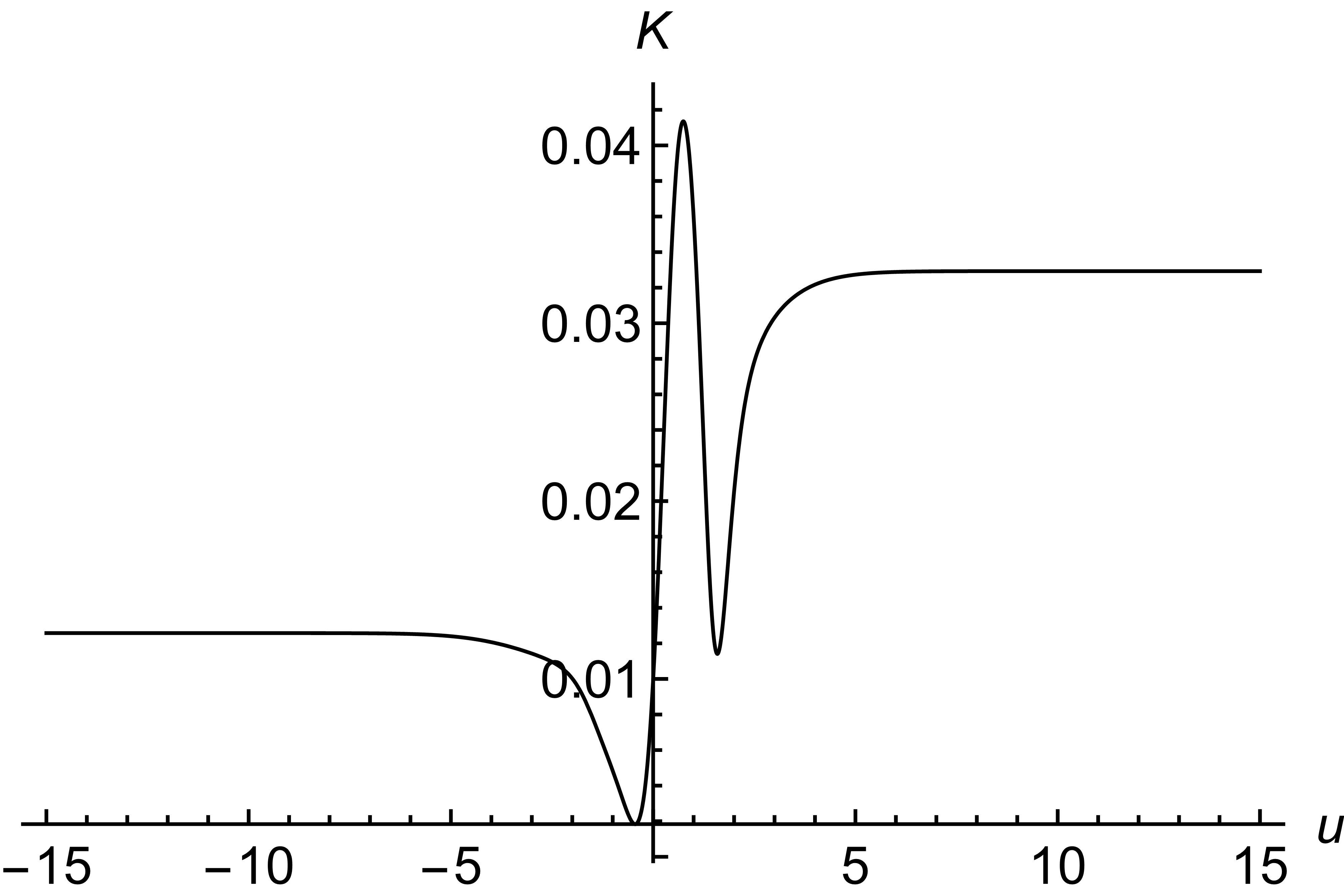}
     \caption{Decreasing of the Kinetic energy for the initial conditions 
          (\ref{24}), and for negative sign of $\dot{\rho_{0}}$.}
          \label{Figure11}
   \end{minipage}
\end{figure}

In the context of this gravitational wave, there also occurs a transfer of 
energy between the wave and the free particle. We observe here that one relevant
feature is the sign of $\dot{\rho_{0}}$. Let us establish the initial conditions

\begin{equation} \label{24}
\rho_{0}=0.6, \ \ \ \phi_{0}=0, \ \ \ z_{0}=0, \ \ \ \dot{\rho_{0}}=\pm 0.2, 
\ \ \ \dot{\phi_{0}}=0, \ \ \ \dot{z_{0}}=0 \ .
\end{equation}
For the initial conditions given above, 
the Kinetic energy of the particle increases or decreases after the passage of 
the wave, according to the 
sign of $\dot{\rho_{0}}$, as shown in Figures [\ref{Figure10}] and 
[\ref{Figure11}].

We recall that the dot represents variation with respect to $u$. Therefore, if
$\dot{\rho_{0}}>0$, the particle is initially approaching the gyratonic source.
Likewise, if $\dot{\rho_{0}}<0$, the particle is initially getting away from the
source. As Figures [\ref{Figure10}] ($\dot{\rho_0}>0$) and [\ref{Figure11}] 
($\dot{\rho_0}<0$) indicate, the particle gain or loose energy as it approaches 
or gets away from the source located around the axis $z=0$.

\begin{figure}[!htb]
   \begin{minipage}{0.49\linewidth}
     \centering
     \includegraphics[width=\textwidth]{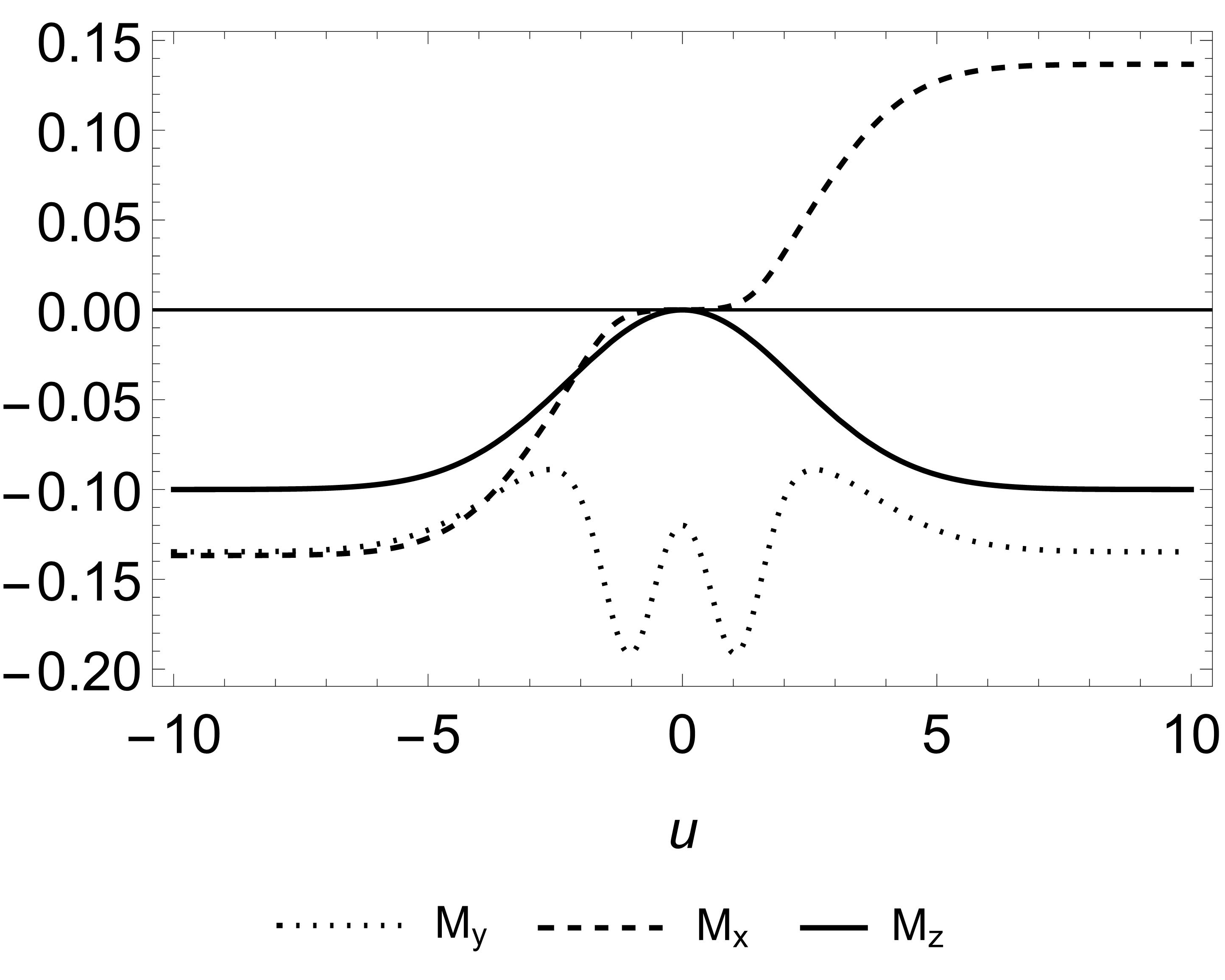}
     \caption{Angular momentum of the particle for the 
     initial conditions (\ref{23}).}\label{Figura12}
   \end{minipage}\hfill
   \begin {minipage}{0.49\linewidth}
     \centering
     \includegraphics[width=\textwidth]{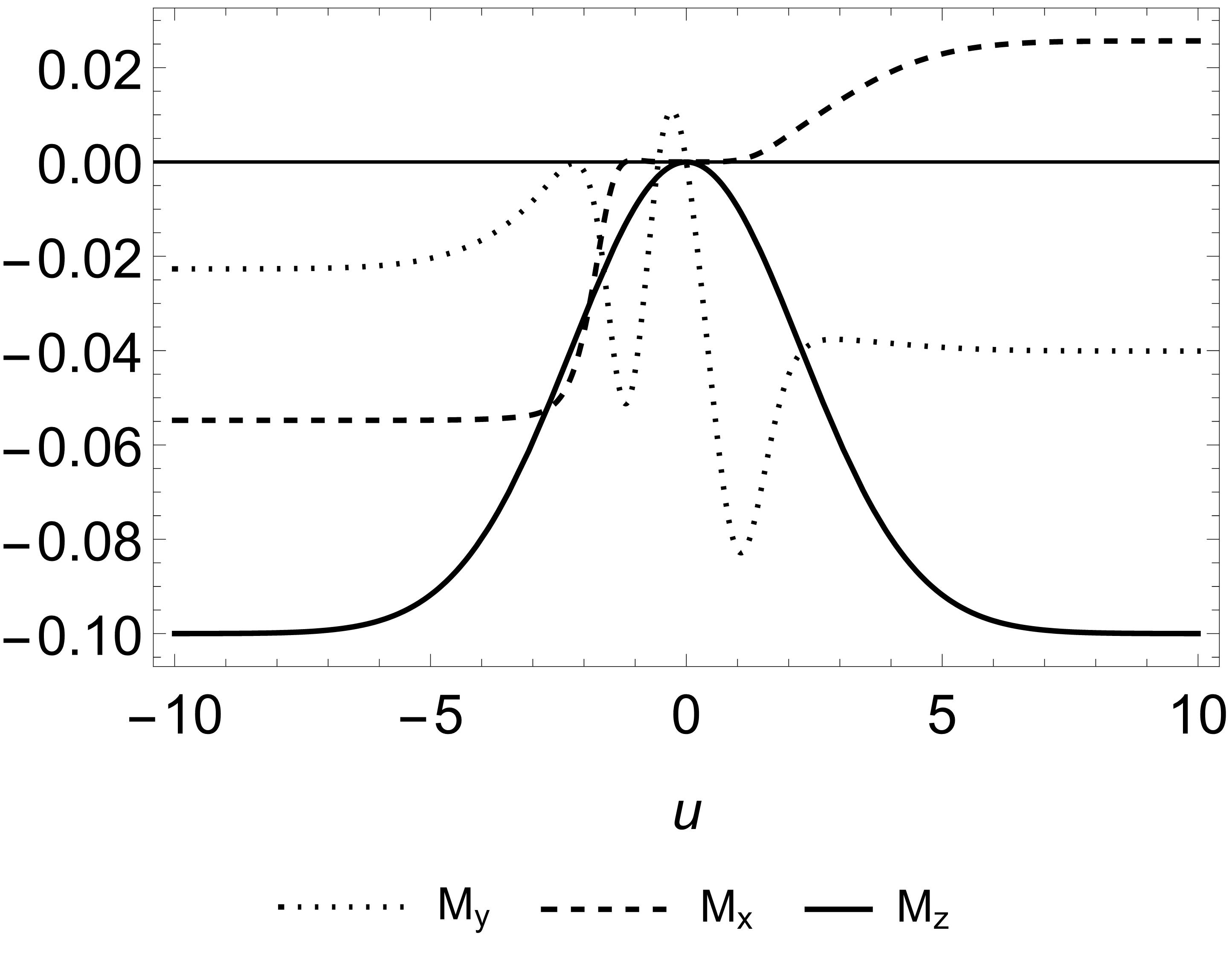}
     \caption{Angular momentum of the particle for the initial conditions
     (\ref{24}), and $\dot{\rho_0}>0$.}\label{Figura13}
   \end{minipage}
	\end{figure}

Still in the context of $H_0$ given (\ref{16}), we display in Figures 
[\ref{Figura12}] and [\ref{Figura13}] the angular momentum per unit mass of the 
particle with respect to $u$. 
The definitions in polar coordinates are 

\begin{eqnarray}
M_{x}&=&\sin{\phi}\left(\rho\dot{z}-z\dot{\rho}\right)-
\rho z\dot{\phi}\cos{\phi}\,, \nonumber \\
M_{y}&=&\cos{\phi}\left(z\dot{\rho}-\rho\dot{z}\right)-
\rho z\dot{\phi}\sin{\phi}\,, \nonumber \\
M_{z}&=&\rho^{2}\dot{\phi}\,.
\label{25}
\end{eqnarray}
Figure [\ref{Figura12}] is constructed taking into
account initial conditions (\ref{23}), whereas in Figure [\ref{Figura13}] we 
consider the initial conditions (\ref{24}), with $\dot{\rho_0}>0$. The two 
Figures clearly show the variation of the angular momentum of the free particle
caused by the gravitational wave. The quantity $M^2=M_x^2 +M_y^2+M_z^2$ is
conserved in the context of Figure [\ref{Figura12}], but not in the context of
Figure [\ref{Figura13}]. This feature confirms that there is a transfer of 
angular momentum between the particle and the wave.

\section{Final Comments}

The non-linear pp-waves, as investigated in Refs. \cite{Ehlers,Ehlers-2}, are
simple, elegant and natural vacuum solutions of Einstein's equations. Although
they have been studied for a quite long time, it is reasonable to say that the
full physical consequences of the non-linear waves have not been investigated so
far. These waves share similarities with plane monochromatic electromagnetic
waves, that are source-free solutions of Maxwell's equations, and that yield an
excellent description of electromagnetic radiation. Ehlers and Kundt showed 
that the principle of linear superposition holds to a certain extent for 
pp-waves, since one may add the amplitudes of two pp-waves that travel in the
same direction (Ref. \cite{Ehlers}, section 2-5.3). This fact supports the
analogy between pp-waves and electromagnetic waves. Gyratonic pp-waves, on the 
other hand, are generated by a source that travels at the speed of light. 
Physically, these waves might be a realistic manifestation of the field
equations. There is no physical principle or physical property that
is violated by the existence of these waves. Therefore, they are legitimate 
manifestations of the theory of general relativity.

We have shown in this article that the Kinetic energy and the angular momentum
per unit mass 
(also the linear momentum, but it has not been explicitly addressed here) of a
free particle vary during the passage of a wave. In particular, the final 
Kinetic energy of a free particle might be smaller than the initial Kinetic
energy. It seems clear that there is a local transfer of 
energy-momentum and angular momentum between the particle and the gravitational 
field of a wave. As a consequence, the wave may gain or loose energy as it 
travels in space. Ehlers and Kundt (Ref. \cite{Ehlers}, section 2-5.8) 
investigated the action of a pp-wave in the form
of a pulse on a cloud of dust particles at rest, and concluded that the 
particles acquire velocities after the pulse has swept the particles. They argue
that this fact suggests convincingly that a cloud or particles  is able to 
extract energy from a gravitational wave, as supported by Bondi \cite{Bondi}.
However, this argument applies when the free particles are initially at rest.
We have seen in this article that if the free particles are not initially at
rest, then the cloud of particles may transfer (positive) energy to the 
gravitational wave.

One interesting extension of the present analysis is the consideration of 
impulsive gravitational waves, as carried out, for instance, in Refs.
\cite{Podolsky-2,Saemann,Podolsky-3}. The profile of an impulsive wave is 
obtained from a single gaussian function, which can be normalised to 1 according
to the expression,

$${1\over {\sigma \sqrt{2\pi}}}\int_{-\infty}^{+\infty} du \;
e^{-{(u^2/ 2\sigma^2})} =1\;,$$
and taking the limit $\sigma \rightarrow 0$. This limit cannot be taken  
in the present analysis, because we cannot introduce a dimensional amplitude in
Eqs. (\ref{14}, \ref{16}) (the second derivative of the gaussian is already
of dimension (length)$^{ -2}$), and also because the second derivative of the 
gaussian cannot be normalised (the integral from $-\infty$ to $+\infty$ of the
latter vanishes). The analysis of the action of impulsive gravitational waves 
on free particles will be carried out elsewhere, and compared to the
results of Refs. \cite{Podolsky-2,Saemann,Podolsky-3}.

The feature discussed in the present article does 
not take place in the context of linearised gravitational waves, in which case
the wave is supposed to be unaffected by the medium, although it becomes 
linearised somehow. Moreover, a gravitational wave, linearised or not, 
imparts a memory effect to the detector, which must involve  a transfer of 
energy-momentum and/or angular momentum to the detector. This issue,
specifically in the context of linearised gravitational waves, must be 
investigated and satisfactorily explained on theoretical grounds.

\end{document}